\documentclass[twocolumn,twoside,slac_two]{revtex4}
\usepackage{graphicx}
\usepackage{fancyhdr}
\begin{document}
\title{Development of a micro-satellite TSUBAME for X-ray polarimetry of GRBs}

%

\author{Shin Kurita, Haruka Ohuchi, Makoto Arimoto, Yoichi Yatsu, Nobuyuki Kawai}
\affiliation{Dept. of Physics, Tokyo Institute of Technology, 2-12-1 Ookayama, Meguro-ku, Tokyo, Japan 152-8551}
\author{Kei Ohta, Masaya Koga, EuGene Kim, Kyosuke Tawara, Souta Suzuki, Kazuyoshi Miyasato, Takashi Nagasu, Shouta Kawajiri, Masanori Matsushita}
\affiliation{Dept. of Mechanical and Aerospace Engineering, Tokyo Institute of Technology, 2-12-1 Ookayama, Meguro-ku, Tokyo, Japan 152-8551}
\author{Saburo Matunaga}
\affiliation{Dept. of Mechanical and Aerospace Engineering, Tokyo Institute of Technology, 2-12-1 Ookayama, Meguro-ku, Tokyo, Japan 152-8551}
\affiliation{Dept. of Space Structure and Materials, Institute of Space and Astronautical Science, Japan 252-5210}
\author{Nagahisa Moriyama, Shin'ichi Kimura}
\affiliation{Dept. of Electrical Engineering, Tokyo University of Science, Japan}
\author{and TSUBAME team}

\begin{abstract}
 TSUBAME is a micro-satellite that the students of Tokyo Institute of Technology took the lead to develop for measuring hard X-ray polarization of Gamma-Ray Bursts(GRBs) in order to reveal the nature of the central engine of GRBs.
 TSUBAME has two instruments: Wide-field Burst Monitor (WBM) and Hard X-ray Compton Polarimeter (HXCP).
 We aim to start observing with HXCP in 15 seconds by pointing the spacecraft using Control Moment Gyro. 
 
 In August 2014, we assembled TSUBAME and performed an integration test during ~2 weeks.
 On Nov 6 2014, TSUBAME was launched from Russia and it was put into Sun-synchronous orbit at 500 km above the ground.
 However, serious trouble occurred to the ham radio equipment.
 Therefore we could not start up the X-ray sensors until Feb 10 2015.
 In this paper, we report the system of TSUBAME and the progress after the launch.
\end{abstract}

\maketitle

\thispagestyle{fancy}


 \section{Introduction}
 Gamma-ray bursts(GRBs) are known as one of the powerful explosions in the universe.
 However, it is not yet clear how to form the collimated outflow wave and accelerate particles.
 In recent theoretical models, the magnetic field plays a important role that connects the central engine and the relativistic outflow.
 The magnetic field might affect the charged particles in the shock front that radiates the X-ray prompt emission.
 Therefore, the X-ray polarization information of prompt emission can constrain the physical process that generate the relativistic outflow from GRBs.
 
\begin{figure}[h!]
 \centering
 \includegraphics[width=60mm]{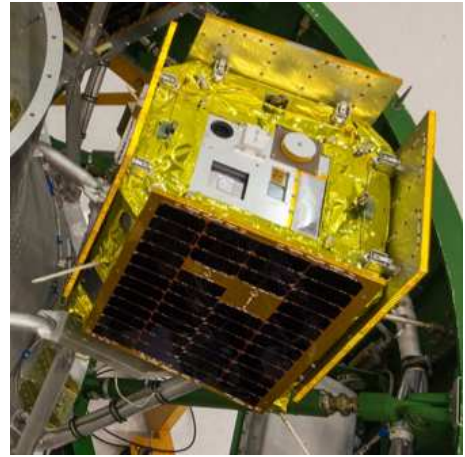}
 \caption{Overview of a micro-satellite TSUBAME.} \label{fig:tsubame_overview}
\end{figure}

 To reveal the  physical process of GRBs, we developed a micro-satellite TSUBAME (Figure.\ref{fig:tsubame_overview}).
 The details of TSUBAME's BUS system are summarized in Table.\ref{tab:bus_system_of_tsubame}.
 
\begin{table*}[tbhp]
 \begin{center}
  \begin{tabular}{|l|l|l|}
   \hline
   Size & \multicolumn{2}{|l|}{610 x 610 x 540 mm$^3$} \\
   \hline
   Mass & \multicolumn{2}{|l|}{48.6 kg}\\
   \hline
   Orbit & \multicolumn{2}{|l|}{500 km (Sun Synchronous)} \\
   \hline
   Launch & \multicolumn{2}{|l|}{Nov 6, 2014}\\
   \hline
   Electrical Power Supply & Cell & InGaP/InGaAs/Ge \\
   \cline{2-3}
   & Power & 130 W (EOL)\\
   \cline{2-3}
   & Battery & 360Wh (Li-Polymer) \\
   \hline
   Command \&  & Tx & S-band (BSPK-100 kbps)\\
   Data Handling(CDH) & & UHF (CW/GMSK-9600bps AFSK-1200bps)\\
   \cline{2-3}
   & Rx & VHF AFSK-1200bps\\
   \hline
   Attitude & \multicolumn{2}{|l|}{3-axis bias momentum control}\\
   \cline{2-3}
   Determination and& Actuator & Control Moment Gyro./Magnetic Torquers\\
   \cline{2-3}
   Control Systems (ADCS) & Sensor & Gyro (MEMS/FOG), Sun Sensor, Magnetometer, Star Tracker, GPS\\
   \hline
  \end{tabular}
  \caption{BUS System of TSUBAME.}
  \label{tab:bus_system_of_tsubame}
 \end{center}
\end{table*}

 \section{GAMMA-RAY OBSERVATION SYSTEM}
  \subsection{Overview}
\begin{figure}[h!]
 \centering
 \includegraphics[width=60mm]{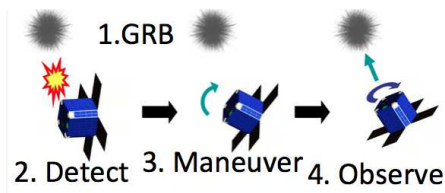}
 \caption{Sequence when a GRB occurs.} \label{fig:maneuver}
\end{figure}

  The micro-satellite TSUBAME has 2 instruments for X-ray polarimetry of GRBs:
  Wide-field Burst Monitor(WBM) and Hard X-ray Compton Polarimeter(HXCP).
  The HXCP utilizes the asymmetry of the Compton scattering arising from the polarization of incident photons.
  The WBM, which consists of five X-ray detectors mounted on the five faces of the satellite, detects GRBs and determines those positions by comparing the count rates in them.
  To detect GRBs, WBM is always monitoring X-ray from half of the sky.
  If WBM detects a GRB, this satellite points to the target rapidly using a high-speed attitude control system and starts the observation within 15 s after the detection(Figure.\ref{fig:maneuver}).

  \subsection{Hard X-ray Compton Polarimeter}
  To maximize the cross section of the Compton scattering, HXCP employed 8 $\times$ 8 ch plastic scintillators connected with the multi-anode photo-multipliers(MAPMTs).
  These scintillators are surrounded by 28 ch CsI scintillators connected with 5 $\times$ 5 mm$^2$ APDs.
  Incident photons are scattered at plastic scintillators, and absorbed at CsI scintillators.
  Linearly polarized photons tend to be scattered perpendicular to the polarization plane.
  HXCP measures the X-ray polarization using this angular dependence.
  
\begin{figure*}[t]
	\begin{center}	
	 \begin{tabular}{c}
	  
	  \begin{minipage}{0.49\hsize}
	   \includegraphics[width=80mm]{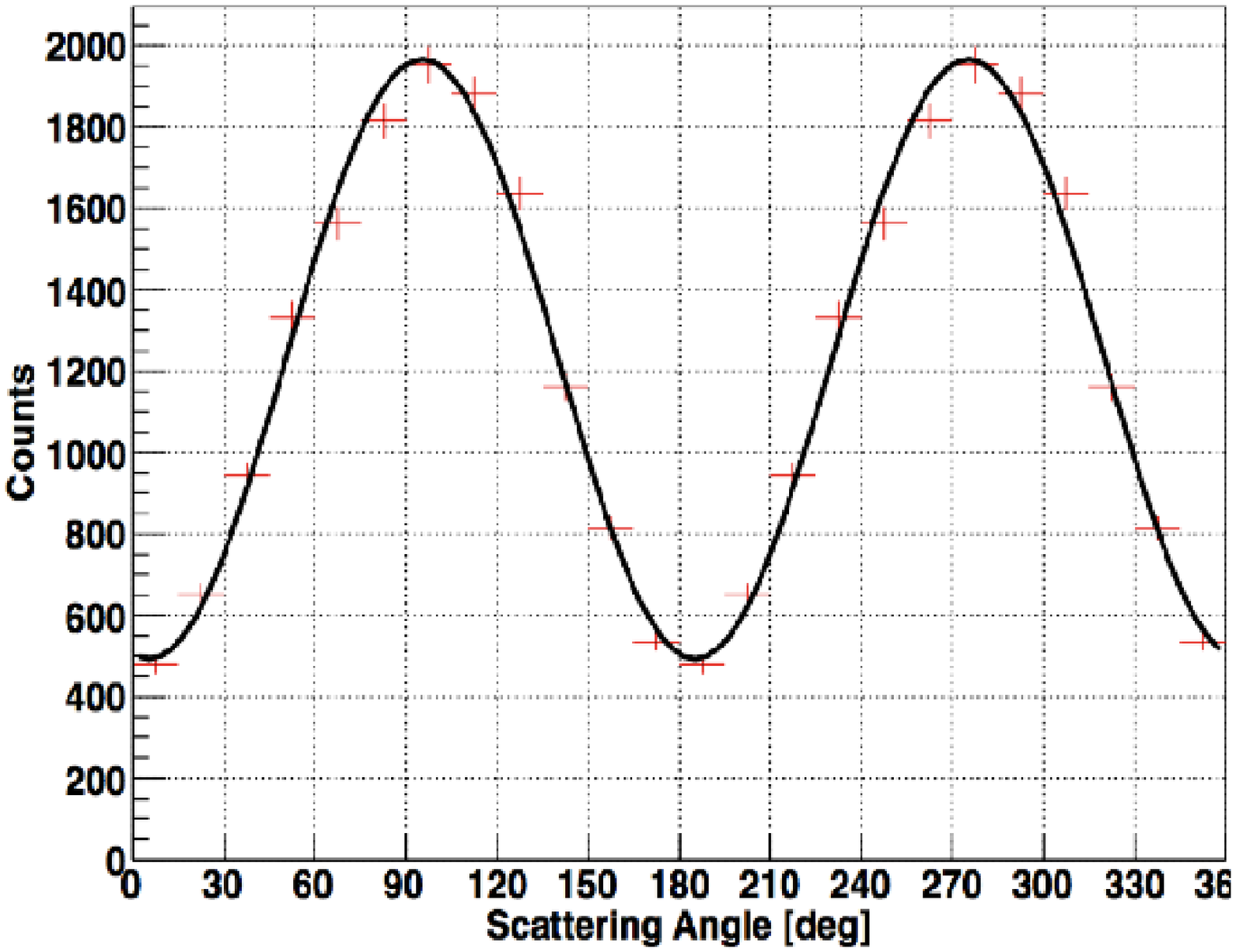}
	   \caption{Modulation Curve with 80 \% on-axis 90 \% synchrotron X-ray.
	   We got 68.5 $\pm$ 0.3 \% of modulation factor.}
	   \label{fig:modulation_0deg}
	  \end{minipage}
	  
	  \begin{minipage}{0.49\hsize}
	   \includegraphics[width=80mm]{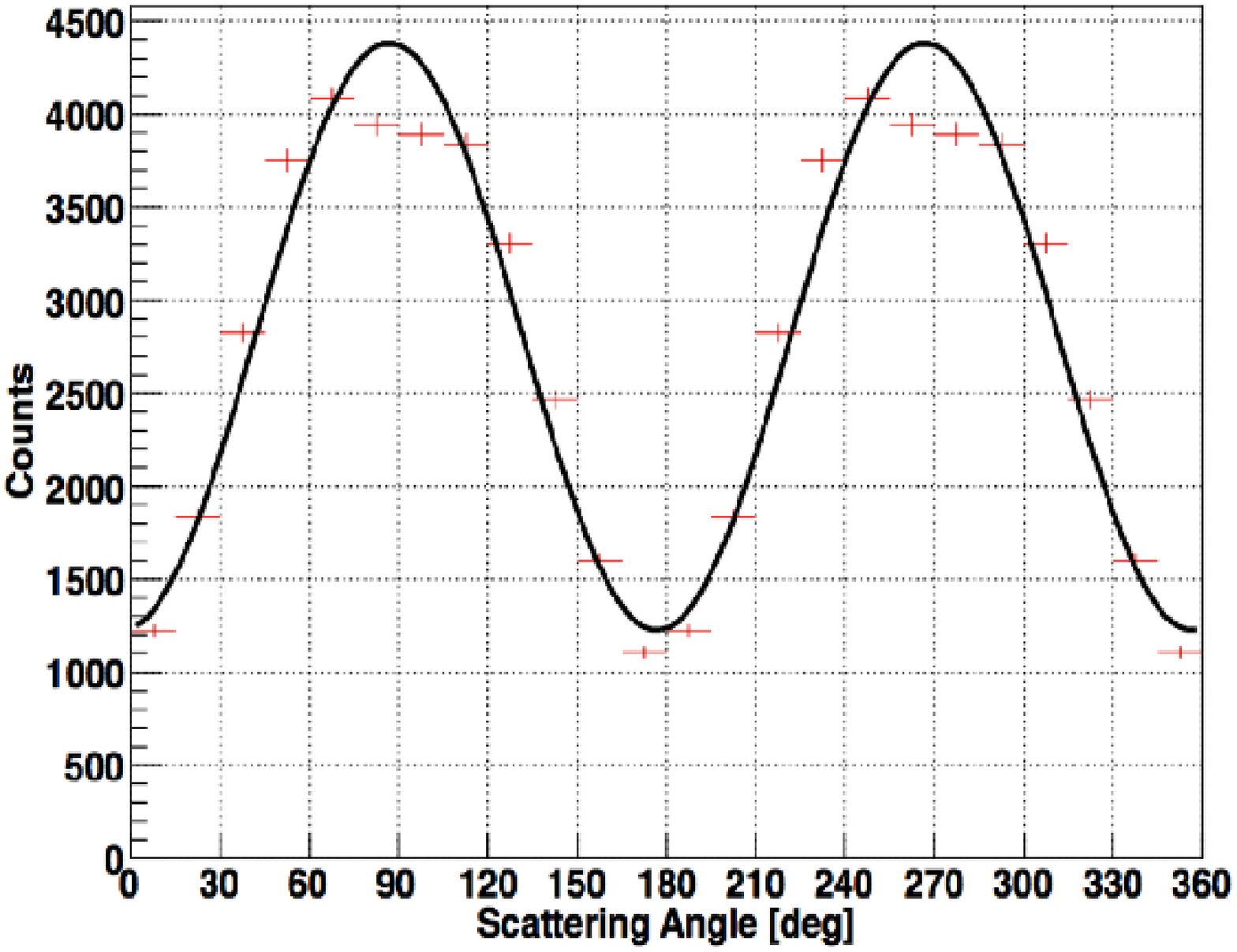}
	   \caption{Modulation Curve with 80 \% off-axis 90 \% synchrotron X-ray.
	   The incident angle is 30 degree.
	   We got 62.6 $\pm$ 0.6 \% of modulation factor.}
	   \label{fig:modulation_30deg}
	  \end{minipage}
	 
	 \end{tabular}
	\end{center}
  \end{figure*}

  In order to evaluate and demonstrate the performance of the HXCP, we executed the performance test at a synchrotron beam facility, the Photon Factory of KEK the high energy accelerator research organization, in December 2012.
  We irradiated 80 \% on-axis 90 \% polarized synchrotron X-ray.
  Before calculating the azimuthal scattering angle for every single photon, we filtered the accumulated photon events with three criteria: (1) a lower threshold energy for PMTs, $>$ 2keV, (2) a lower threshold energy for APDs, $>$ 20keV, (3) a geometric constraint on the energy distribution between the recoil electron and the scattered photon.
  The modulation curve as functions of the azimuthal scattering angle using the screened event data with the above three criteria are shown in Figure.\ref{fig:modulation_0deg}.
  We obtained 68 \% of the modulation factor.
  As well, we drew a modulation curve using the photon event date of same X-ray with off-axis, and we obtained over 60 \% of the modulation factor(Figure.\ref{fig:modulation_30deg}).

  \begin{figure*}[t]
   \centering
   \includegraphics[width=135mm]{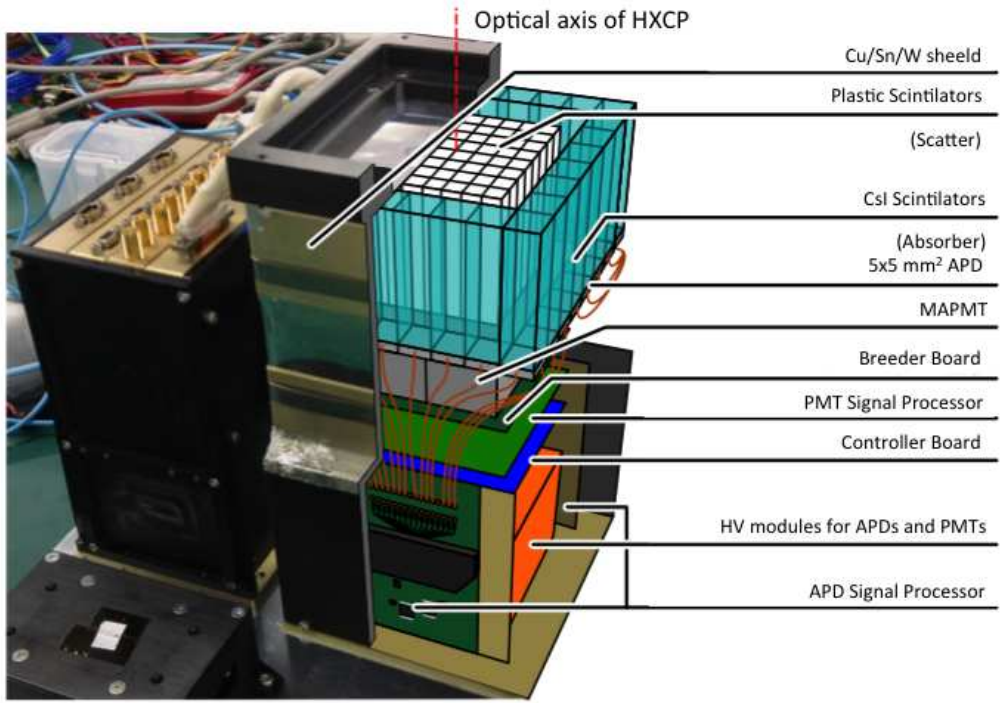}
   \caption{Overview of HXCP and a circuit box. }
   \label{fig:hxcp_overview}
  \end{figure*}
  
  \subsection{Wide-field of Burst Monitor}
  Wide-field Burst Monitor(WBM) consists of five sintillation gamma-ray counters with APDs and CsI scintillator for real time detection and localization of GRBs.
  These detectors are mounted on the five faces of TSUBAME.

\begin{figure*}[t]
   \begin{center}
	\begin{tabular}[t]{c}
	 
	\begin{minipage}{0.49\hsize}
	 \includegraphics[width=80mm]{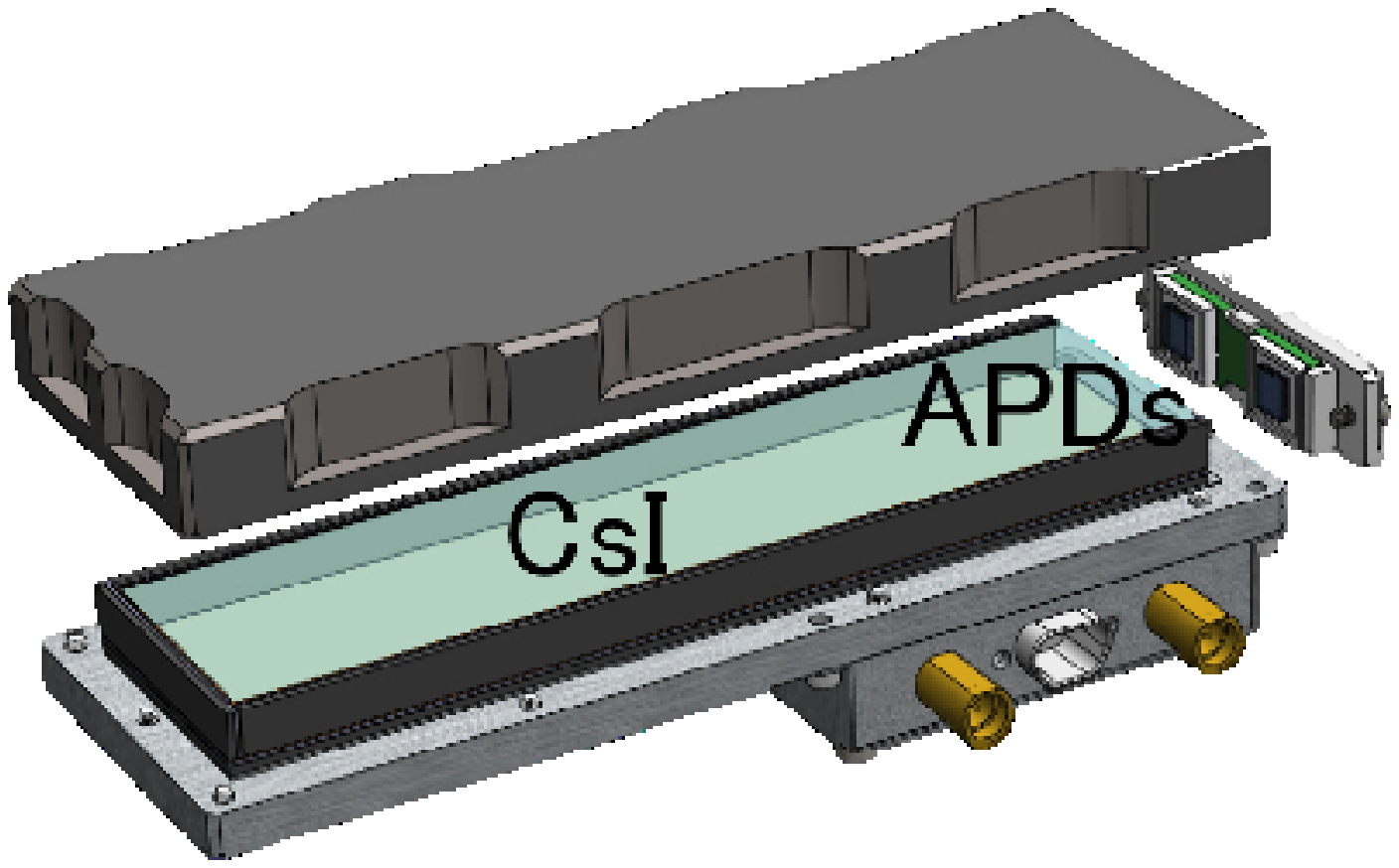}
	 \caption{Overview of WBM.}
	 \label{fig:wbm_model}
	\end{minipage}
	 
	 \begin{minipage}{0.49\hsize}
	  \includegraphics[width=80mm]{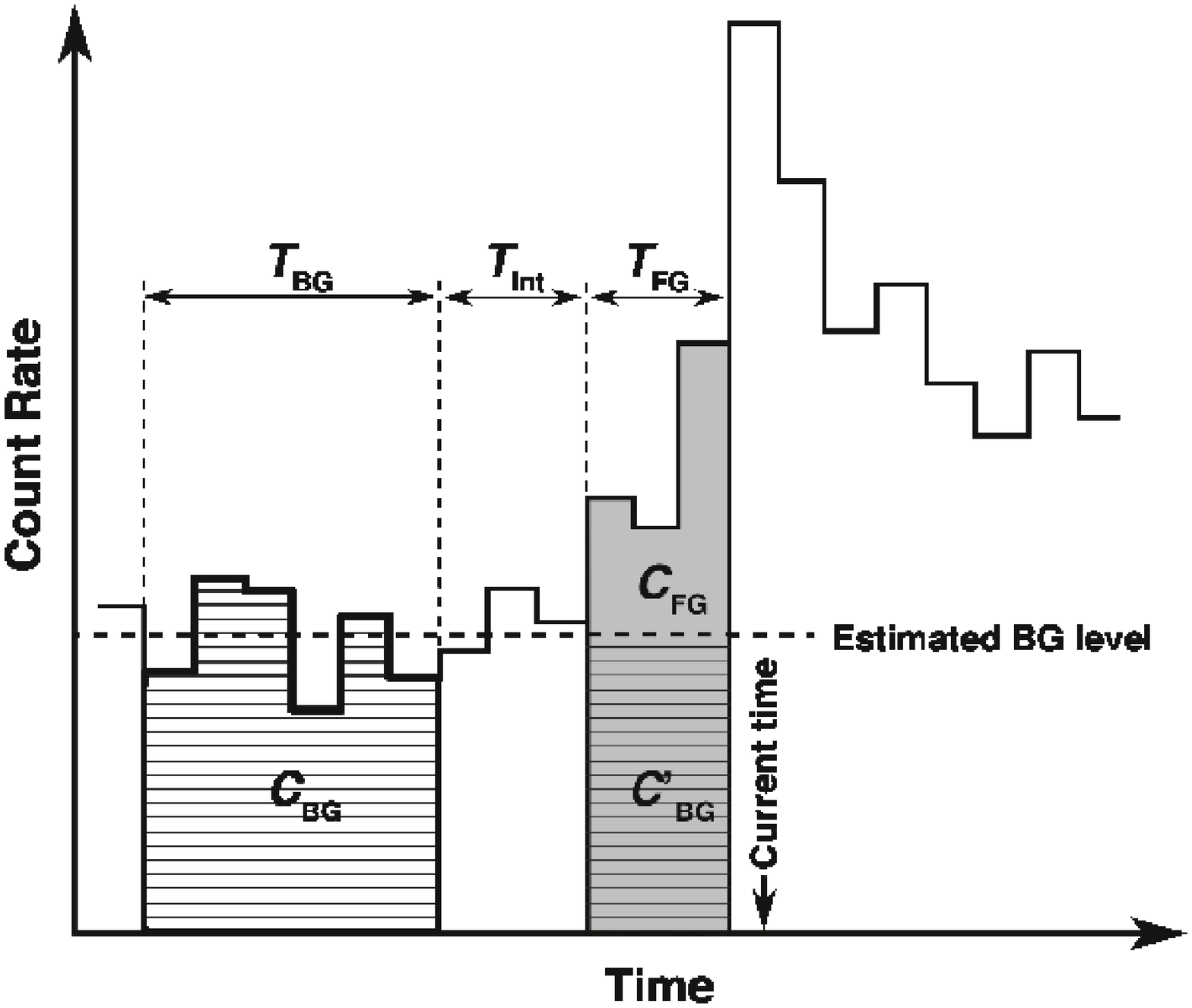}
	  \caption{Description of the WBM trigger algorithm. }
	  \label{fig:trigger_algorithm}
	 \end{minipage}
	 
	\end{tabular}
   \end{center}
  \end{figure*}

  The time scale of GRBs ranges from milli-second to kilo-seconds.
  In order to start a pointing observation with the HXCP as soon as possible, WBM must detect any GRBs faster.
  To detect GRBs, on-board CPU checks the variation of X-ray count rate of WBM every 125 msec.
  WBM employs 4 trigger systems with different time constants which covers short GRBs and long GRBs.
  In designing these trigger systems, we refer to the trigger systems of HETE-II(Figure.\ref{fig:trigger_algorithm}).
  The on-board CPU estimates the background events, $C'_{\rm BG}$, at the time window, $T_{\rm FG}$, from the past background events, $C_{\rm BG}$, at the time window, $T_{\rm BG}$, using the recorded light curves.
  The net photon events, $C_{\rm Sig}$, can be written as
  \begin{equation}
   C_{\rm Sig} = C_{\rm FG} - C'_{\rm BG} = C_{\rm FG} - \frac{T_{\rm FG}}{T_{\rm BG}}.
  \end{equation}
  
  If we assume that the background rate fluctuates statistically, the significance, $s$, of the $C_{\rm Sig}$ comparing with the background fluctuation is expressed as $s = C_{\rm Sig}/\sqrt{C_{\rm BG}}$.
  The on-board CPU evaluates $s^2$ at every 125 ms to check the variability of the light curves for each energy band and detector.
  
  When trigger system detect a GRB, WBM determines the position of GRB by comparing the event rates of these five X-ray detectors with an accurary of 5 degree for a bright GRB with an occurrence rate of 10 bursts per year.
  This method have been employed by the BATSE abroad CGRO and the GBM on the Fermi gamma-ray observatory.
  However the expected position accuracy is not so good, this accuracy is sufficient for polarimetry using the flight model HXCP with a wide FoV.
  
  After WBM determines the position of GRB, TSUBAME starts high-speed attitude control with Control Moment Gyroscopes pointing to the GRB.
  Using Control Moment Gyroscopes, TSUBAME can maneuver ten times as fast as Swift.
  
 \section{Operation}
\begin{figure*}[t]
  \centering
  \includegraphics[width=150mm]{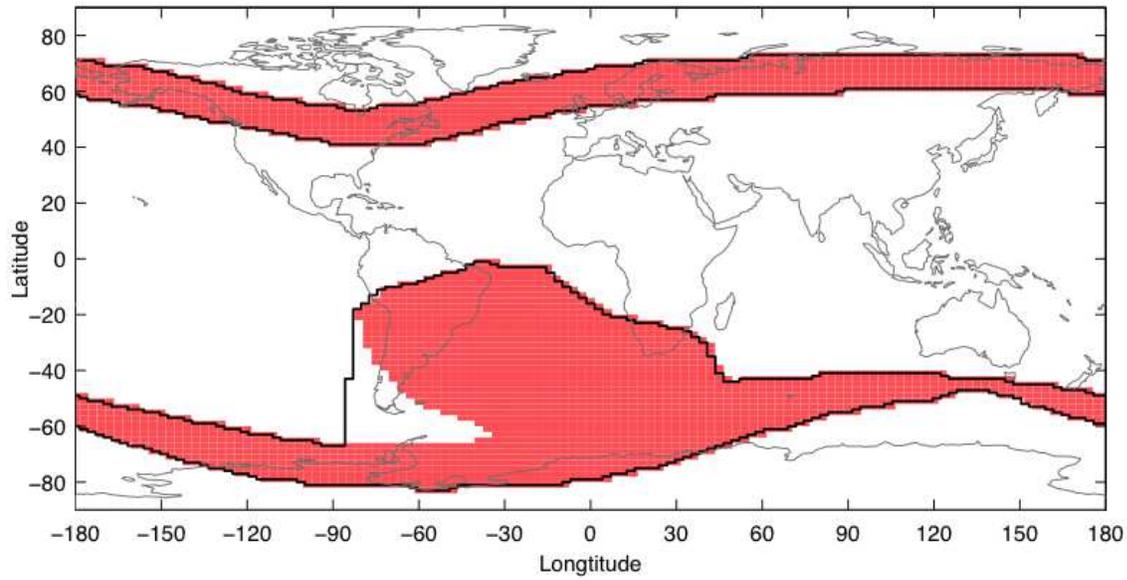}
  \caption{Stored SAA map.}
  \label{fig:saa_map}
 \end{figure*}

  To protect HXCP from the damages in South Atlantic Anomaly(SAA) and Auroral zone, we will have to drop the high voltage of MAPMTs.
  TSUBAME has three ways to judge whether it is in SAA or not.
  At first, on-board CPU compares GPS location to the stored SAA map in Figure.\ref{fig:saa_map}.
  The shading area describes the prohibited regions in which the particle flux of electrons above 1 MeV exceeds 300 counts s$^{-1}$ cm$^{-2}$ and that of protons above 20 MeV exceeds 0.3 counts s$^{-1}$ cm$^{-2}$.
  This map is based on Space Environment Information System and Cute-1.7+APDII data in orbit.
  Secondly, we set specific GRB trigger patterns as indications of SAA.
  When TSUBAME enters this area, the count rate of WBM increase slowly than GRBs.
  Because the trigger system of WBM can detect every increase of count rate depending on parameters, we set 1 trigger for SAA trigger.
  Finally, we predict the timing from the satellite orbital element and uplink the stored commands via 430MHz band.
  
  However the trigger system detects the variation of count rate of WBM, it can not distinguish GRBs or known blight sources.
  In order to avoid false triggers caused by a steady bright X-ray sources appearing from the horizon, GRB triggers must be disabled by the stored commands at these predicted timings.
  
\begin{figure*}[t]
  \centering
  \includegraphics[width=150mm]{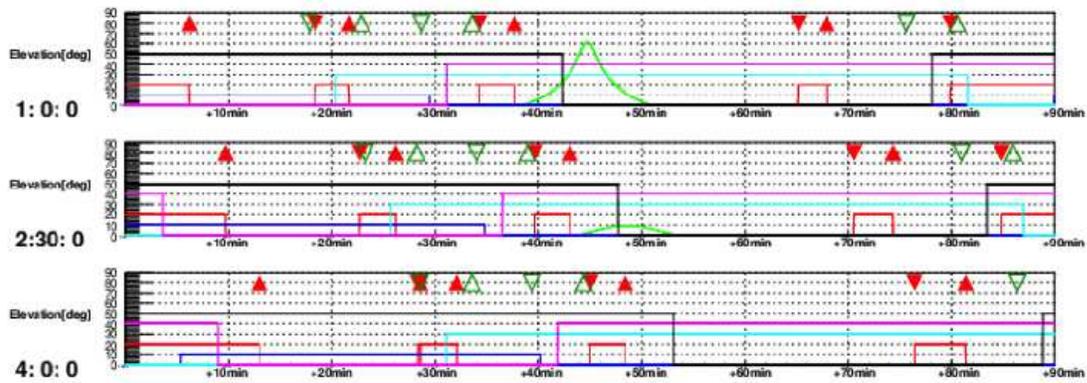}
  \caption{Graphical check sheet generated by the dp10 of TSUBAME.
  The green line shows the elevation of TSUBAME.
  The red, blue, magenta, cyan, and black lines shows the flag of SAA, shade, Sco-X1, Cygnus-X1, and Crab respectively.
  The red triangles show the timings of HV on/off.
  The green triangles show the timings of WBM trigger enable/disable.
  }
  \label{fig:dp10}
 \end{figure*}

  For above reasons, we must uplink over-100 commands for an observation.
  In order to execute commands at relevant timings, we developed the command generating program.
  Refering to DP10 which is a graphical command check program used in Suzaku operation, we developed the program, the dp10 of TSUBAME, to display the types and timings of commands at a time axis so as to confirm generated commands by human eyes.
  Figure.\ref{fig:dp10} is a graphical check sheet generated by the dp10 of TSUBAME.
  
 After observations, we downlink the observation data via S-band.
 The maximum data size is about 6 MByte and it will take 80 min to downlink.
 \\
 \\
 
 \section{After Launch}
 In August 2014, we assembled TSUBAME and performed an integration test. 
 We operated TSUBAME continuously during $\sim$ 2 weeks.
 We successfully operated all the sequence.
 
 TSUBAME was launched from Russia on Nov 6 2014.
 After the launch, we succeeded to obtain telemetries at Tokyo Tech ground station.
 We confirmed that TSUBAME had succeeded to expand the solar cell panels and to point at the Sun,
 However, serious trouble occurred to the ham radio equipment.
 Therefore we could not start up the X-ray sensors until Feb 10 2015.


\begin{thebibliography}{9}
	\bibitem{Yatsu2014} Y. Yatsu et al, ``Pre-flight performance of a micro-satellite TSUBAME for X-ray polatimetry of gamma-ray bursts'', Proceeding of SPIE, Vol.9144, 2014
	\bibitem{Tavenner2003} T. Tavenner et al, ``The Effectiveness of the HETE-2 Triggering Algorithm'', American Institute of Physics Conference Services 662, 2003
   \end{thebibliography}
\end{document}